\documentclass[aps,superscriptaddress,showpacs,nofootinbib, 10pt,notitlepage]{revtex4-1}
\usepackage{mathrsfs}
\usepackage{bigints}
\usepackage{latexsym,bm}
\usepackage{graphicx}
\usepackage{indentfirst}
\usepackage{slashed}
\usepackage{amssymb}
\usepackage{amsmath}
\usepackage{bbm}
\usepackage{color}
\usepackage[dvipsnames]{xcolor}
\usepackage{epsfig}
\usepackage{multirow}%
\usepackage{rotating}
\usepackage{epstopdf}
\usepackage{extarrows}
\usepackage{tabularx}
\usepackage{soul} 
\usepackage{xcolor}
\usepackage{lipsum}
\usepackage[colorlinks=true,allcolors=BlueViolet,hyperfootnotes=false]{hyperref}

\usepackage[utf8]{inputenc}

\usepackage[inline]{enumitem}

\newcommand{\be}{\begin{equation}} \newcommand{\ee}{\end{equation}}
\newcommand{\ba}{\begin{array}{c}} \newcommand{\ea}{\end{array}}
\newcommand{\bea}{\begin{eqnarray}} \newcommand{\eea}{\end{eqnarray}}

\newcommand{\Tcc}{T_{cc}^+}

\newcommand{\oinvb}{\mathcal{O}(\beta^{-1})}

\newcommand{\ogbbB}{\mathcal{O}\left(\frac{\gamma_b}{\beta^2}\right)}
\newcommand{\aLO}{a_\text{LO}}
\newcommand{\rLO}{r_\text{LO}}
\newcommand{\aNLO}{a_\text{NLO}}
\newcommand{\rNLO}{r_\text{NLO}}
\newcommand{\ie}{\textit{i.e.}}
\newcommand{\eg}{\textit{e.g.}~}
\newcommand{\cf}{\textit{cf.}}
\newcommand{\LZdr}{\mathcal{L}(Z, \delta r)}
\newcommand{\dr}{\delta r}
\newcommand{\Dsz}{D^{\ast}_{s0}(2317)^\pm}

\usepackage{accents}

\begin{document}
\title{Compositeness of S-wave weakly-bound states \\ from next-to-leading order Weinberg's relations}

\newcommand{\ific}{Instituto de F\'{\i}sica Corpuscular (centro mixto CSIC-UV),
Institutos de Investigaci\'on de Paterna,
C/Catedr\'atico Jos\'e Beltr\'an 2, E-46980 Paterna, Valencia, Spain}

\author{M.~Albaladejo}
\email{Miguel.Albaladejo@ific.uv.es}
\author{J.~Nieves}
\email{jmnieves@ific.uv.es}
\affiliation{\ific}

\date{\today}

\definecolor{citecolor}{rgb}{0.15,0.15,0.60}

\begin{abstract}
We discuss a model-independent estimator of the likelihood of the compositeness  of a shallow S-wave bound or virtual state. The approach is based on an extension of Weinberg's relations in \textcolor{citecolor}{Phys.~Rev.~\textbf{137}, B672 (1965)} \cite{Weinberg:1965zz} and it relies only on the proximity of the energy of the state to the two-hadron threshold to which it significantly couples. The scheme only makes use of  the experimental scattering length and the effective range low energy parameters, and it is shown to be fully consistent for predominantly molecular hadrons. As explicit applications, we analyse the case of the deuteron, the $^1{\rm S}_0$ nucleon-nucleon virtual state and the exotic $\Dsz$, and find strong support to the molecular interpretation in all cases. Results are less conclusive for the $\Dsz$, since the binding energy of this state is significantly higher than that of the deuteron, and the approach employed here is  at the limit of its applicability. We also qualitatively address the case of the recently discovered $\Tcc$ state, within the isospin limit to avoid the complexity of the very close thresholds $D^0D^{\ast +}$ and $D^+D^{\ast 0}$, which could mask the ingredients of the approach proposed in this work.
\end{abstract}

\maketitle

\section{Introduction} \label{sec:intro}

Quantum Chromodynamics (QCD), the theory of strong interactions, generates a rich spectrum of hadrons, most of which can be classified according to simple constituent quark models  \cite{Gell-Mann:1964ewy,ZweigOne,ZweigTwo,Godfrey:1985xj,Capstick:1986ter} as $q\bar{q}$ (mesons) and $qqq$ (baryons) states. Despite this fact, the last two decades have witnessed the discovery of many states that defy this simple classification \cite{ParticleDataGroup:2020ssz}. Different worldwide experiments (BaBar, Belle, BES, LHCb, $\ldots$) have reported the observation of a plethora of unstable states and peaks  in mass-distributions located surprisingly close to different two bottomed/charmed-hadron thresholds, such as the $XYZ$ mesons, the $P_c$ pentaquarks~\cite{Belle:2003nnu,Belle:2011aa,BESIII:2013ris,Belle:2013yex,BESIII:2013mhi,BESIII:2013ouc,LHCb:2015yax,LHCb:2019kea,BESIII:2020qkh}, or the doubly charmed $T_{cc}^+$ state \cite{LHCb:2021auc,LHCb:2021vvq}. There exist also clear examples of exotic candidates in the open charm and bottom sectors, \eg $D^{\ast}_{s0}(2317)$, $D^{\ast}_{0}(2300)$, $\Lambda_c(2595)$ \cite{BaBar:2003oey,Belle:2003nsh,CLEO:1994oxm}, $\ldots$,  $B_1(5721)$,  $B_2^*(5747)$, $\Xi_b(6227)$~\cite{D0:2007vzd,CDF:2013www,LHCb:2015aaf, LHCb:2018vuc, LHCb:2020xpu}, etc. These states are often interpreted as hadron molecules~\cite{Barnes:2003dj,vanBeveren:2003kd,Kolomeitsev:2003ac,Szczepaniak:2003vy,Voloshin:2003nt,Close:2003sg,Swanson:2003tb,Tornqvist:2004qy,Guo:2006fu,Gamermann:2006nm,Flynn:2007ki,Faessler:2007gv,Lutz:2007sk, Dong:2009yp,Gamermann:2009fv,Wu:2010jy,Bondar:2011ev,Wang:2011rga,Yang:2011wz,Garcia-Recio:2008rjt,Romanets:2012hm,Garcia-Recio:2012lts,Wu:2012md,Liu:2012zya,Garcia-Recio:2013gaa,Wang:2013cya,Guo:2013sya,Aceti:2014uea,Albaladejo:2015lob,Albaladejo:2016jsg,Albaladejo:2016lbb,Nieves:2017jjx,Du:2017zvv,Nieves:2019nol,Wang:2018jlv,Karliner:2015ina,Xiao:2013yca,Nieves:2019jhp,Liu:2019tjn,Du:2019pij,Liu:2019stu,Lu:2014ina,Dong:2021bvy,Du:2021fmf,Du:2021zzh,Dong:2021juy,Guo:2017jvc,Guo:2020vmu,Feijoo:2021ppq,Albaladejo:2021vln}, or compact tetraquarks/pentaquarks~\cite{Esposito:2016noz,Maiani:2004vq,Maiani:2014aja,Braaten:2013boa,Dias:2013xfa,Maiani:2014aja,Qiao:2013raa,Deng:2014gqa,Ali:2011ug,Cheng:2003kg,Terasaki:2003qa,Dmitrasinovic:2005gc,Bracco:2005kt,Wang:2006uba,Chen:2004dy,Kim:2022mpa,Maiani:2015vwa,Lebed:2015tna,Li:2015gta,Ghosh:2015xqp,Wang:2015epa,Zhu:2015bba,Richard:2017una,Hiyama:2018ukv,Barnes:2003vb}, and when allowed by their quantum numbers and flavour content, there are also attempts to describe these states as predominant $q\bar{q}$ or $qqq$  structures~\cite{Colangelo:2003vg,Cahn:2003cw,Browder:2003fk,Godfrey:2003kg,Narison:2003td,Bardeen:2003kt,Nowak:2003ra,Browder:2003fk,Dai:2003yg,Lee:2004gt,Lakhina:2006fy,Segovia:2013wma,Yoshida:2015tia,Ortega:2016mms}. Other possibilities (hybrids, virtual poles, etc.) for the nature of these exotics~\cite{Close:2003mb,Li:2004sta,Wang:2006bs,Dubynskiy:2008mq,Voloshin:2013dpa,Fernandez-Ramirez:2019koa,Pilloni:2016obd} or the role of kinematic (non-dynamical) effects (chiefly triangle singularities) in the interpretation of the observed peaks have also been stressed~\cite{Wang:2013hga,Chen:2013coa,Swanson:2014tra,Guo:2014iya,Guo:2015umn,Liu:2015fea,Albaladejo:2015lob,Szczepaniak:2015eza,HALQCD:2016ofq,Pilloni:2016obd,Bayar:2016ftu,Guo:2019twa}. Further discussions and references can be found in Refs.~\cite{Lebed:2016hpi,Chen:2016spr,Ali:2017jda,Karliner:2017qhf,Olsen:2017bmm,Brambilla:2019esw}. In addition to knowing how many of these states exist and their masses and widths, it is clearly a fundamental task in hadron physics to study the dynamical details of their structure. Such analysis will be invaluable in improving our understanding of strong interactions.

It is a direct consequence of unitarity that the inverse single-channel two-particle scattering amplitude $f$
is given in terms of the phase shift $\delta$ by $f(E)^{-1} = k \cot\delta(k) -ik$, with $k=\sqrt{2\mu E}$, for non-relativistic kinematics, $\mu$  the reduced mass of the scattering particles, and $E$ the energy of the system relative to the threshold $(m_1+m_2)$. The real part of the inverse scattering amplitude is a polynomial  in even powers of $k$ and, in the case of S-wave, it leads to the effective range expansion (ERE):
\begin{equation}
k \cot\delta(k) = \frac{1}{a}+\frac12 r k^2+ {\cal O}\left(k^4\right)~, \label{eq:ERE}
\end{equation}
where the parameters $a$ and $r$ are called the scattering length and effective range, respectively. Weinberg's compositeness rules~\cite{Weinberg:1965zz} connect these low energy observables with the probability $Z$ that an S-wave shallow bound state is found in a bare elementary-particle state, which might allow to interpret $X=(1-Z)$ as a kind of molecular probability:
\begin{subequations}
\label{eq:Wein}
\begin{align}
a  & =\underbrace{-\frac{2}{\gamma_b} \left(\frac{1-Z}{2-Z} \right)}_{\aLO}\ +\ \oinvb~, \label{eq:Wein_a}\\
r  & =\underbrace{-\frac{1}{\gamma_b}\left(\frac{Z}{1-Z}\right)}_{\rLO}\ +\ \oinvb~, \label{eq:Wein_r}
\end{align}
\end{subequations}
where $\gamma_b=\sqrt{2\mu |E_b|}$ (with $E_b<0$, the binding energy), $\beta$ denotes the next momentum scale that is not treated explicitly in the ERE, with $1/\beta$  providing an estimate for the interaction range corrections. The work of Ref.~\cite{Weinberg:1965zz} showed that the experimental values for $a$ and $r$ from $pn$ scattering  give strong model-independent evidence that the deuteron is composite, \ie, the probability $Z$ of ending the deuteron in a bare elementary particle state is very small. However, as commonly acknowledged (see \eg Refs.~\cite{Matuschek:2020gqe, Esposito:2021vhu,Li:2021cue,Song:2022yvz}), this does not follow from the naive evaluation of $X(a,r)=1/\sqrt{1+2r/a}$, easily derived from Eqs.~\eqref{eq:Wein}, which gives the meaningless result of $X=1.68>1$ for a probability. Indeed, the above formula and this numerical value for $X$ are not given in Ref.~\cite{Weinberg:1965zz}. As explicitly pointed out by Weinberg, the key token for the deuteron compositeness is the fact that $r$ is small and positive of the order of the range $\sim m_\pi^{-1}$ of the $pn$ interaction, rather than large and negative. This discussion clearly shows the ambiguities affecting any conclusion about the nature of an exotic state based uniquely on a blind numerical computation of $X$, neglecting $\oinvb$ corrections. Different applications, re-derivations, re-interpretation and extensions of Weinberg's compositeness relations have been proposed~\cite{Baru:2003qq,Gamermann:2009uq,Baru:2010ww,Hanhart:2011jz,Aceti:2012dd,Hyodo:2011qc,Sekihara:2014kya,Kamiya:2015aea,Garcia-Recio:2015jsa,Guo:2015daa,Kamiya:2016oao,Sekihara:2016xnq,Oller:2017alp,Matuschek:2020gqe,Esposito:2021vhu,Li:2021cue,Kinugawa:2021ykv,Song:2022yvz,Sazdjian:2022kaf,Bruns:2022hmb,Kinugawa:2022fzn}, discussing or trying to improve upon various aspects of the derivation of the criterion in Ref.~\cite{Weinberg:1965zz}, in order to increase the cases to which it can be applied, \textit{i.e.}, by considering also coupled channels.

We discuss now two subtle points about the interpretation of $Z$.\footnote{Our discussion is similar to that of Ref.~\cite{Guo:2017jvc}.} In Eq.~(18) of Ref.~\cite{Weinberg:1965zz}, $Z$ is defined as a probability between the states of the continuum and a bare, compact state, with the patent difficulty that the latter is not, in general, a physical, asymptotic state. Therefore, the effects produced by the interacting hadron cloud must be taken into account. From an Effective Field Theory point of view, the effects of a bare coupling (of a bare, compact state to a two-hadron continuum state) plus other perturbative hadron--hadron interactions are translated into the physical coupling of the state, $g^2$. The latter, as we will see below [\textit{cf.} Eq.~\eqref{eq:zgt}], enters into the LO term contribution to $Z$, and thus the interpretation of $Z$ in terms of compositeness is unchanged~\cite{Weinberg:1962hj,Weinberg:1963zza} (see also Refs.~\cite{Guo:2017jvc,Albaladejo:2021cxj}). From a different point of view, $Z$ is also a renormalization field factor, and in this sense it is in general a scheme-dependent quantity. However, it is crucial that in the weak binding limit $(\gamma_b \ll \beta)$, the LO contribution to $Z$ is determined by the pole position and the residue of the two-hadron scattering amplitude at the physical pole~\cite{Weinberg:1965zz,Matuschek:2020gqe}. This term is non-analytical in $E$ and it is thus scheme-independent. Indeed, the LO contribution is written in terms of two measurable quantities (position and coupling of the bound state), and this model-independent contribution to $Z$ becomes a measure of the compositeness. The scheme-dependent terms of $Z$, for instance those analytic in $E$, need to be fixed by some renormalization condition, but importantly they are suppressed by a factor of the order $\mathcal{O}(\gamma_b/\beta)$~\cite{Weinberg:1965zz,Matuschek:2020gqe}. Indeed, in the weak binding limit one has:
\begin{equation}
\left(1-Z\right)  = \frac{\mu g^2}{\gamma_b}+{\cal O}\left(\gamma_b/\beta\right), \qquad g^2 \equiv \lim_{E-E_b} (E_b-E)f(E) \label{eq:zgt}   
\end{equation}
The above equation illustrates how the coupling $g^2$ and the binding momentum $\gamma_b$ provide the leading contribution to the probability $X=1-Z$, and that they do not fully determine the sub-leading  $\mathcal{O}(\gamma_b/\beta)$ contributions to $Z$. In addition, we note that Eq.~\eqref{eq:zgt} is consistent with Eqs.~\eqref{eq:Wein} and the approximate relation $\mu g^2 \approx \gamma_b/(1-r\gamma_b)$, valid for loosely bound states. 

In this work, we will delve in the above discussion, and we will present an estimator of the compositeness of a state relying only on the proximity of its mass to a two-hadron threshold to which it significantly couples. Our conclusions will apply only to weakly bound or virtual states since model-independent statements are possible only if $\gamma_b \ll \beta$. The internal structure of a non-shallow bound state cannot be studied without having a knowledge of its wave-function more detailed than what can be inferred from the experimental value of just a few scattering parameters. Hence, one cannot extract model-independent conclusions on the compositeness of a particle located far from the relevant threshold, since one would necessarily have to rely on certain interaction, re-summation and renormalization models. In this respect, the calculation of hadron form factors, which can be accessed by LQCD simulations, can play an important role in determining the internal structure of hadrons~\cite{Flynn:2007ki,Briceno:2020vgp,Albaladejo:2012te}.

For simplicity, we will ignore coupled-channels dynamics, though we will comment in Sec.~\ref{sec:Tcc} on this issue. We will also assume that the particle is stable, otherwise  $Z$ is complex. However, it  might be an adequate approximation to ignore the decay modes of a very narrow resonance.

The paper is organized as follows. After this introduction, we present in Sec.~\ref{sec:formalism} the model-independent estimator of the likelihood of the compositeness of a shallow S-wave bound/virtual state proposed in this work. Next we show explicit applications and analyse the case of the deuteron (Sec.~\ref{sec:deu}), the $^1{\rm S}_0$ nucleon-nucleon virtual state (Sec.~\ref{sec:vir}) and the exotic $\Dsz$   and $\Tcc$ states in  Secs.~\ref{sec:ds0} and \ref{sec:Tcc}, respectively. We collect the most important conclusions of our work in Sec.~\ref{sec:summary}.

\section{Likelihood of the compositeness of a weakly bound state} \label{sec:formalism}

The parameters of the ERE [\cf~Eq.~\eqref{eq:ERE}] depend in turn on $\gamma_b$, which is determined by the binding energy of the state. The large values for both $a$ and $r$ when $Z$ is not zero appear because of the $1/\gamma_b$ contributions in Eqs.~\eqref{eq:Wein}. They may suggest that the next-to-leading order (NLO) approximation to the ERE, consisting in neglecting ${\cal O}\left(k^4\right)$ terms, may itself break down when the particle is elementary. This, however, does not happen since only the first two terms in the expansion of $k \cot\delta(k)$  in powers
of $k^2$  become of order $\gamma_b$ for $Z\ne 0$ and $k\approx \gamma_b$. The third and higher terms are smaller by powers of $\gamma_b/\beta$~\cite{Weinberg:1965zz}. As a consequence, the approximate relation 
\begin{equation}\label{eq:polecond}
\gamma_b \approx -\frac{1}{a}+\frac12 r \gamma_b^2 
\end{equation}
is expected to  be fulfilled with great accuracy for weakly bound states. This is indeed the case for the deuteron. The above relation does not tell anything about the elementarity of the particle, since it follows from the requirement  $\cot\delta(i \gamma_b)=+i$. In fact, it is exactly satisfied by $a_{\rm LO}$ and $r_{\rm LO}$ for all $Z$ [\cf~Eqs.~\eqref{eq:Wein}]. However, the deviations of the actual scattering length and effective range from their $\gamma_b-$expansion leading-order (LO) values $a_{\rm LO}$ and $r_{\rm LO}$ encode some valuable information on the compositeness of the state.  Moreover, from the discussion above, the third and higher terms in the ERE of Eq.~\eqref{eq:ERE} could provide at most corrections\footnote{The point that Weinberg makes in Ref.~\cite{Weinberg:1965zz} is that the coefficient of the $k^4$ term in the ERE does not diverge as $1/\gamma_b^3$ in the weak-binding limit, which will affect to the approximate relation of Eq.~\eqref{eq:polecond} at order $\gamma_b$. However, one cannot discard that this coefficient could scale like $Z/\gamma_b^2$. The factor of $Z$ in front is because this $1/\gamma_b^2$ behaviour will not appear in the absence of compact bare states ($Z=0$). The reasoning runs in parallel for any other higher order term of the ERE. Hence, we simply discard any new contribution of order ${\cal O}(\gamma_b)$ to Eq.~\eqref{eq:polecond}  coming from the ERE terms beyond the scattering length and effective-range.} of order ${\cal O}( Z\gamma_b^2/\beta)$ to the difference $\gamma_b-\left(- 1/a+r \gamma_b^2/2\right)$.  With these ideas in mind, we introduce a phenomenological term $\delta r$  to estimate the NLO contribution to the effective range $r$, within its expansion in powers of the binding momentum $\gamma_b$,

\begin{subequations}\label{eq:arNLO}
\begin{equation}
r = \underbrace{-\frac{1}{\gamma_b}\left(\frac{Z}{1-Z}\right)+\dr}_{\rNLO}+ \ogbbB, \quad \rNLO = \rLO + \dr~, \label{eq:rNLO}
\end{equation}
This correction, $\dr$, is expected to be of the order of the range of the interaction [${\cal O}(1/\beta)$]. The scattering length will have a similar NLO contribution, $\delta a$, on top of the LO term in Eq.~\eqref{eq:Wein_a}.  We fix  this analogous NLO contribution to the scattering length such  that the difference $\gamma_b-\left( -1/a_{\rm NLO}+r_{\rm NLO} \gamma_b^2/2\right)$ [\cf~Eq.~\eqref{eq:polecond}] deviates from zero in terms of the order ${\cal O}\left(\gamma_b^3/\beta^2\right)$. This is to say, we require that the $\gamma_b^2$ term in the Taylor expansion (in powers of $\gamma_b$) of the quantity $\left[\gamma_b-\left( -1/a_{\rm NLO}+r_{\rm NLO} \gamma_b^2/2\right)\right]$ vanishes. In this way, we  obtain:
\begin{equation}
a =  \underbrace{-\frac{2}{\gamma_b} \left(\frac{1-Z}{2-Z} \right)- \frac{\delta r}{2} \left(\frac{1-Z}{1-\frac{Z}{2}}\right)^2}_{\aNLO} + \ogbbB,\, \quad \aNLO = \aLO - \frac{\delta r}{2} \left(\frac{1-Z}{1-\frac{Z}{2}}\right)^2. \label{eq:aNLO}
\end{equation}
\end{subequations}
The relations given in Eqs.~\eqref{eq:arNLO} constitute the main result of this work. The key point is that the same parameter $\delta r$ appear in both $a_\text{NLO}$ and $r_\text{NLO}$. Therefore, they provide a model-independent scheme to correlate the NLO corrections to $a$ and $r$, which turns out to be consistent as long as $Z \simeq 0$, of the order of $ \mathcal{O} \left(\gamma_b/\beta \ll 1\right)$ or smaller. The fact that $Z$  is required to be at most of order $\mathcal{O} \left(\gamma_b/\beta \right)$   is because the ${\cal O}(k^4)$ terms in the ERE expansion could lead to corrections of order $\mathcal{O} \left(Z\gamma_b^2/\beta\right)$ in Eq.~\eqref{eq:polecond}, as mentioned before. These corrections should be, at most, comparable to those of ${\cal O}\left(\gamma_b^3/\beta^2\right)$ neglected in the present scheme to correlate $\rNLO$ and $\aNLO$. When $Z$ takes these small values, it would mean that the model-independent contribution from the coupling $g^2$ in Eq.~\eqref{eq:zgt} should be large, giving a strong support to the molecular nature of the weakly bound state. In other words, a shallow bound state which couples to a two-hadron system, for which the scattering length and effective-range could be accurately described by $\aNLO$ and $\rNLO$ with values of $Z$ of the order of $\left(\gamma_b/\beta\right)$, small but not necessarily zero, can be reasonably seen as a hadron molecule. This is to say, its low energy properties, including the binding, can be naturally accommodated as result of the two-hadron interaction. To study systems for which $Z$ could take larger values,\footnote{That is, larger than $\mathcal{O}(\gamma_b/\beta)$ but not necessarily $Z \to 1$. For this latter case, corresponding to a purely compact state, the scattering amplitude that arises by considering a bare state propagator is exactly the ERE truncated at $\mathcal{O}(k^2)$. Therefore, in the $Z \to 1$ case, no additional terms in the ERE beyond $a$ and $r$.} the ${\cal O}(k^4)$ or higher terms in the ERE expansion will be required, which points to  the need in these cases to include additional details  of the  short distance dynamics.

Given the experimental values for the mass, or equivalently the binding momentum $\gamma_b^\text{exp}$, the associated scattering length ($a_{\rm exp}$), and the effective range ($r_{\rm exp}$) parameters of a two-particle weakly bound state, we propose to study the following two dimensional (2D) distribution:
\begin{align}
    \LZdr = & \frac{1}{3} \left[ \left(\frac{a_{\rm exp}-a_{\rm NLO}}{\Delta a_{\rm exp}}\right)^2 + \left(\frac{r_{\rm exp}-r_{\rm NLO}}{\Delta r_{\rm exp}}\right)^2 + \left(\frac{\gamma^{\rm exp}_b-\gamma^{\rm NLO}_b}{\Delta \gamma_b^{\rm exp}}\right)^2 \right],  \label{eq:liHood}
\end{align}
to estimate the likelihood of the compositeness of the state. Above, $\gamma^{\rm NLO}_b$ is given by 
\begin{equation}
\gamma^\text{NLO}_b = \frac{1 - \sqrt{1 + 2 \rNLO /\aNLO}}{\rNLO}~,
\end{equation}
which exactly satisfies $\left[\gamma^{\rm NLO}_b-\left( -1/a_{\rm NLO}+r_{\rm NLO} (\gamma^{\rm NLO}_b)^2/2\right)\right]=0$. Consistent with the order at which we are working, we evaluate $a_{\rm NLO}$ and $r_{\rm NLO}$  using $\gamma^{\rm exp}_b$. It is important to notice that $\Delta \gamma_b^{\rm exp}, \Delta a_{\rm exp}$, and $\Delta r_{\rm exp}$ should be fixed taking into account not only the uncertainties on the determinations of these observables, but also the expected accuracy of their NLO approximation, \textit{i.e.}, an $\mathcal{O}(\gamma_b^2/\beta^2)$ relative error. If the actual experimental errors are smaller than this expected accuracy, then a relative error $(\gamma_b/\beta)^2$ should be taken instead.

\begin{figure*}\centering
\includegraphics[scale=0.99]{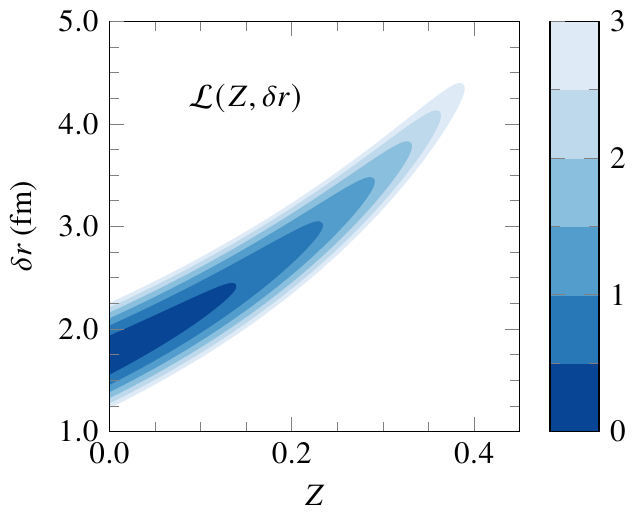}
\includegraphics[scale=0.99]{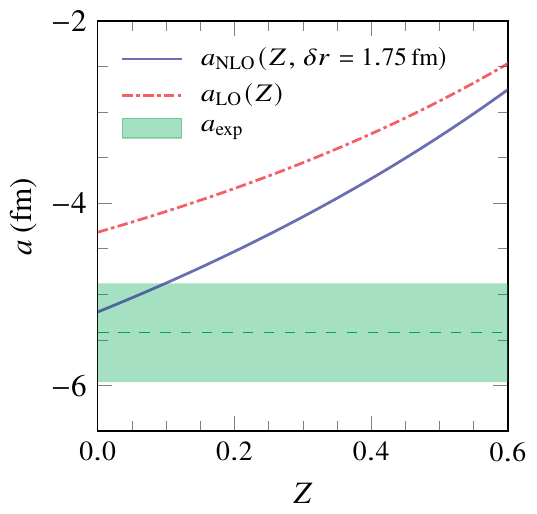}
\includegraphics[scale=0.99]{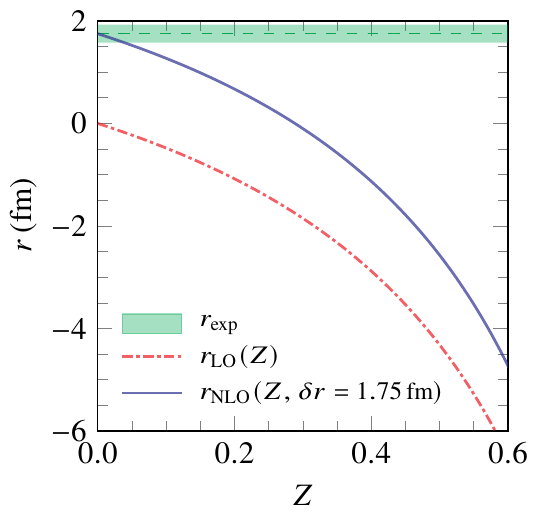}
\caption{Left: Distribution of Eq.~\eqref{eq:liHood} for different values of the probability $Z$ of ending the deuteron in a bare elementary particle}, as a function of the NLO contribution $\delta r$ [Eq.~\eqref{eq:rNLO}] to the effective range of the $pn$ interaction in the isoscalar $^3{\rm S}_1$ wave. Middle and right: $Z$ dependence of $a$ (middle) and $r$ (right) at LO (red, dashed-dotted lines) and NLO (blue, solid lines), compared with the experimental values (green bands).\label{fig:deuteron}
\end{figure*}

\section{The deuteron} \label{sec:deu}

We first apply the compositeness distribution of Eq.~\eqref{eq:liHood} to the paradigmatic case of the deuteron, whose properties are known very precisely: $E_b^{\rm exp}=-2.224575 (9)$ MeV [or equivalently $\gamma_b^{\rm exp}=0.2316068 (5)\, {\rm fm}^{-1}$] \cite{VanDerLeun:1982bhg}, and $a_{\rm exp}=-5.42 (1)$ fm and $r_{\rm exp}=1.75 (1)$ fm from the Granada-group  analysis of the $pn$ isoscalar $^3{\rm S}_1$  wave \cite{NavarroPerez:2014ovp} (see also Ref.~\cite{RuizArriola:2019nnv}). The analysis of the latter work  includes statistical errors, stemming from the data uncertainties for a fixed form of the potential, and systematic errors arising from the different most-likely forms of the potentials. Assuming they are independent, the total uncertainty corresponds to adding both errors in quadrature. Despite including systematic uncertainties, the errors of  $a$ and $r$ turn out to be much smaller than the accuracy that can be expected from the NLO approximation, $(\gamma_b^2/m_\pi^2)\sim 10\%$, taking $\beta \simeq m_\pi$. Therefore, instead of taking the small errors quoted above, we fix $\Delta \gamma^{\rm exp}_b$, $\Delta a_{\rm exp}$, and $\Delta r_{\rm exp}$ in Eq.~\eqref{eq:liHood} assuming a relative error of 10\%. With all these inputs, we show in Fig.~\ref{fig:deuteron} the 2D distribution ${\cal L}(Z, \delta r)$ for  the case of the deuteron. It strongly supports molecular probabilities $(1-Z)$ quite close to one, in conjunction with values of the NLO $\delta r$ contribution of the order of $1/m_\pi \sim 1.4\,\text{fm}$, as expected. We also see in Fig.~\ref{fig:deuteron} (middle and right plots) that for a value $\dr = 1.75\,\text{fm}$,  $\aNLO$ and $\rNLO$ are  closer to the experimental values than the corresponding LO predictions [\cf~Eq.~\eqref{eq:Wein}]. This model-independent analysis provides a consistent picture, where values of $Z$ greater than $\gamma_b^{\rm exp}/m_\pi \sim 0.3$ are very implausible, which confirms the dominant molecular structure of the deuteron \cite{Weinberg:1965zz}. 

\section{\boldmath The $\Dsz$} \label{sec:ds0}

This scalar narrow resonance ($\Gamma < 3.8$ MeV), which lies $45\,\text{MeV}$ below the $DK$ threshold and  has valence-quark  content $c\bar s$, was discovered in 2003 by the BaBar Collaboration \cite{BaBar:2003oey}. Its abnormally light mass cannot be easily accommodated within constituent quark models~\cite{Fayyazuddin:2003aa,Sadzikowski:2003jy, Lakhina:2006fy,Segovia:2013wma, Ortega:2016mms}, and as a consequence, it is common to describe this exotic resonance as the result of the S-wave $DK$ interaction~\cite{Guo:2006fu,Flynn:2007ki,Guo:2008gp,Liu:2012zya,Mohler:2013rwa,Lang:2014yfa,MartinezTorres:2014kpc, Bali:2017pdv, Guo:2017jvc,Albaladejo:2018mhb}. Unitarized  heavy-meson chiral approaches predict that the $\Dsz$ would belong to a light-flavor SU(3) anti-triplet, completed by an isospin doublet associated to the scalar $D^*_{0}(2300)$ resonance~\cite{Albaladejo:2016lbb, Du:2017zvv}. Here, we will discuss, within the scheme outlined above, the $DK$ molecular probability of the  $\Dsz$, neglecting the isospin-violating $D_s\pi$ decay channel. Given the lack of $DK$ scattering data, we take the values of the isoscalar S-wave $DK$ scattering length and effective range obtained in Ref.~\cite{MartinezTorres:2014kpc} from the finite volume QCD levels reported in Refs.~\cite{Mohler:2013rwa, Lang:2014yfa}. Namely, we use $a_{\rm exp}=-1.3 (5)\,\text{fm}$ and $r_{\rm exp}=-0.1 (3)\,\text{fm}$. In  addition, we take $E_b^{\rm exp}=-45(4)\,\text{MeV}$, estimated from the experimental masses compiled in the PDG \cite{ParticleDataGroup:2020ssz}, which leads to $\gamma_b^{\rm exp}=0.95 (4)\,\text{fm}^{-1}$ when isospin averaged masses are used for the kaon and $D$ mesons. Due to the large uncertainties affecting both $a_{\rm exp}$  and $r_{\rm exp}$, it is not necessary to take into account the subtleties associated with  the $D^0K^+$--$D^+K^0$ isospin breaking effects. The  scale $\beta$ is in this case  of the order of $300\,\text{MeV}$, estimated from the expected effects induced by the nearest  $D_s\eta$ threshold \cite{Matuschek:2020gqe} and/or by the two-pion exchange interaction, none of which are explicitly treated in the ERE. Hence, we expect  the accuracy of the NLO $\gamma_b-$expansion to be of the order of $(\gamma_b^2/\beta^2)\sim 40\%$, which we adopt for $\Delta \gamma^{\rm exp}_b$, while for $\Delta a_{\rm exp}$ and $\Delta r_{\rm exp}$, we use the errors quoted above. 

\begin{figure}\centering
\includegraphics[scale=0.9]{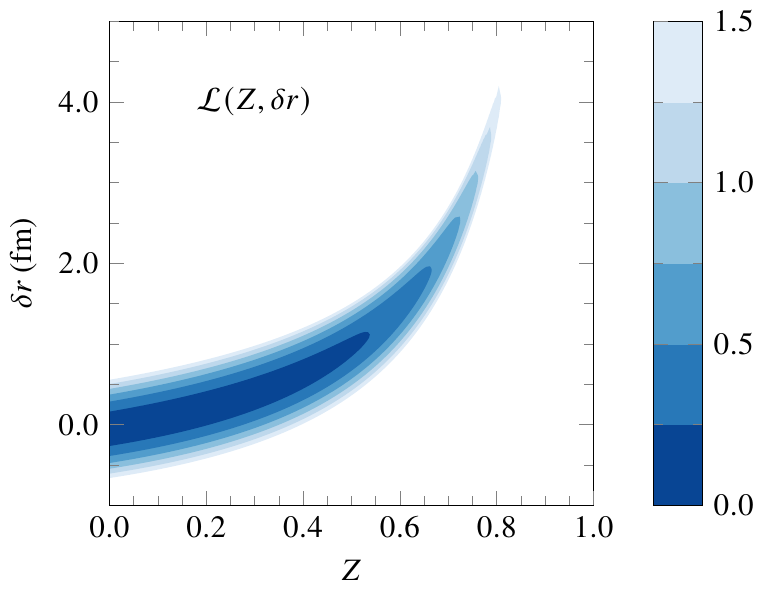}
\caption{Compositeness  distribution for the $\Dsz$.  }\label{fig:ds2317}
\end{figure}

The compositeness 2D distribution of Eq.~\eqref{eq:liHood} for the $\Dsz$  is shown in Fig.~\ref{fig:ds2317}. The results favor $DK$ molecular probabilities of at least 50\%, which is in agreement with previous calculations~\cite{Liu:2012zya, MartinezTorres:2014kpc,Albaladejo:2016hae,Guo:2017jvc,Albaladejo:2018mhb, Matuschek:2020gqe,Song:2022yvz}. We cannot be as predictive in this case as for the deuteron, not only because of the bigger uncertainties of the input, but also because of the larger size of the power-counting parameter $\gamma_b/\beta \sim 0.6$. Nevertheless, the approach is still consistent since values of $Z$ of order ${\cal O}\left(\gamma_b/\beta\right)$, or smaller, are favored by the compositeness distribution for this state. We however should note that the binding energy of the $\Dsz$ is significantly higher than that of deuteron, and the NLO approach employed here is  at the limit of its applicability. To be quantitatively more precise, it would be necessary to know more details about the dynamics at short distances of this exotic state than those encoded in the first two parameters of the ERE.

\section{\boldmath The $^1{\rm S}_0$ nucleon-nucleon virtual state} \label{sec:vir}

The $pn$ scattering length and effective range determined in Ref.~\cite{RuizArriola:2019nnv} for this isovector partial wave are  $a_{\rm exp}=23.735(16)\ \text{fm}$ and $r_{\rm exp}=2.68(3)\ \text{fm}$, respectively, with the total uncertainties obtained by adding statistical and systematic errors in quadrature. This partial wave has a shallow virtual state (pole on the real energy-axis below the threshold on the unphysical sheet), the position of which is determined by the condition $\cot\delta(-i \gamma_v)=+i$, with $\gamma_v=\sqrt{2\mu |E_v|}$ and $E_v<0$, the binding energy of the virtual state. From the ERE, it follows
\begin{equation}
\gamma_v \approx \frac{1}{a}-\frac12 r \gamma_v^2 + {\cal O}(\gamma^4_v)~.\label{eq:effrangevirtual}
\end{equation}
Neglecting ${\cal O}(\gamma^4_v)$ corrections and using  the experimental $a_{\rm exp}$ and $r_{\rm exp}$
values, the above equation leads to
\begin{equation}
   \gamma_v^{\rm exp} \approx \frac{-1 + \sqrt{1 + 2 r_{\rm exp} /a_{\rm exp}}}{r_{\rm exp}} = 0.03999(5)\, {\rm fm}^{-1}~, \label{eq:gvirt}
\end{equation}
where the  error comes from the uncertainties of the experimental ERE parameters. This error turns out to be around a factor of ten greater than the corrections induced by  the ${\cal O}(\gamma^4_v)$ term in Eq.~\eqref{eq:effrangevirtual}. This virtual binding momentum corresponds to $E_v= -0.0663 (4)$ MeV. 

\begin{figure}\centering
\includegraphics[scale=0.9]{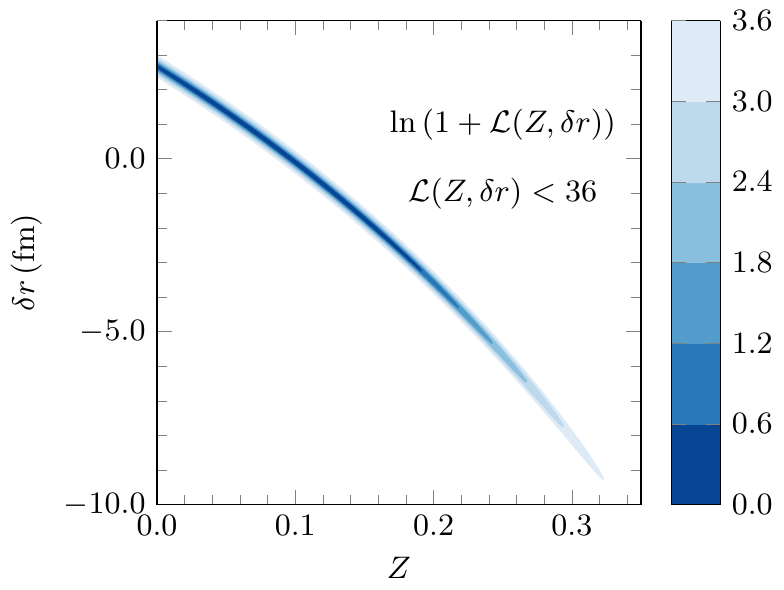}
\caption{Compositeness distribution for  the $^1{\rm S}_0$ nucleon-nucleon virtual state. Note that the $\ln (1+{\cal L})$ function is displayed for ${\cal L}< 36$.} \label{fig:1S0}
\end{figure}

It is not trivial to extend the notion of compositeness to states other than bound states, since wave functions derived from poles on the unphysical sheet are not normalizable and the probabilistic interpretation is lost. They are not QCD asymptotic states, and thus it seems difficult to argue about wave-function components. However, one could think of some variation of QCD parameters, \eg quark masses, such that these virtual states could become physical, bound states. From this perspective, it would make sense to generalize the notion of compositeness. On the other hand, formally relying on the definition of the field renormalization $Z$ in the nonrelativistic theory~\cite{Matuschek:2020gqe}, relations between $a$, $r$ and $Z$ can be derived also for a virtual state with a pole at $k=-i\gamma_v$, and  they are similar to those of a bound state. One should simply  replace $\gamma_b$ with $-\gamma_v$ in $a_{\rm NLO}$ and $r_{\rm NLO}$ given in Eqs.~\eqref{eq:Wein}~and~\eqref{eq:arNLO}. In addition, $\gamma_v^{\rm NLO}$ should be evaluated using Eq.~\eqref{eq:gvirt}, but with the NLO ERE parameters. Thus, we can  use the definition of Eq.~\eqref{eq:liHood} for the compositeness distribution of a virtual state. The accuracy of the NLO $\gamma-$expansion $(\gamma_v^2/m_\pi^2)\sim 0.3\%$ is larger than the experimental errors on $a_{\rm exp}$ and $\gamma_v^{\rm exp}$, and we  take it to set $\Delta \gamma_v^{\rm exp} $ and $\Delta a_{\rm exp}$, while we use the experimental error to fix $\Delta r_{\rm exp}$. Due to the high precision of the input parameters, very small variations of $Z$ and $\delta r$ produce quite large changes of ${\cal L}(Z, \delta r)$. In Fig.~\ref{fig:1S0}, we show the neperian logarithm of the compositeness  distribution for this case. We see that for $Z > 0.2$, the 2D function  ${\cal L}(Z, \delta r)$ takes values larger than 2, while its minimum values are found for $Z < 0.1$.  This is consistent with the general understanding that virtual states are the result of two-particle interactions, and in this sense can be considered of molecular type~\cite{Matuschek:2020gqe}.

\begin{figure*}\centering
\includegraphics[scale=0.99]{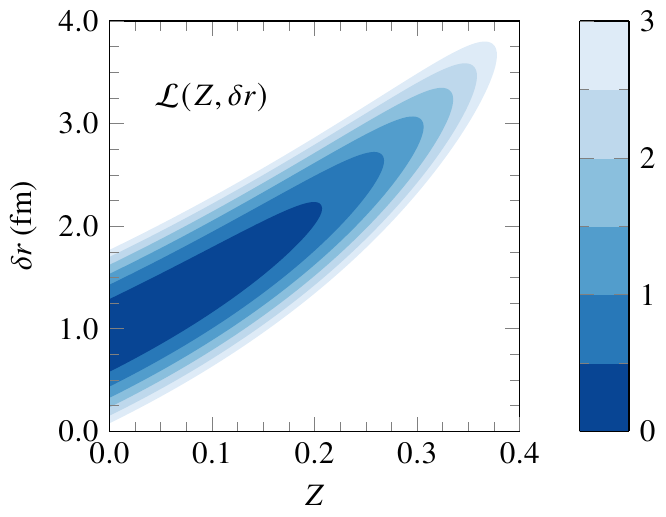}
\includegraphics[scale=0.99]{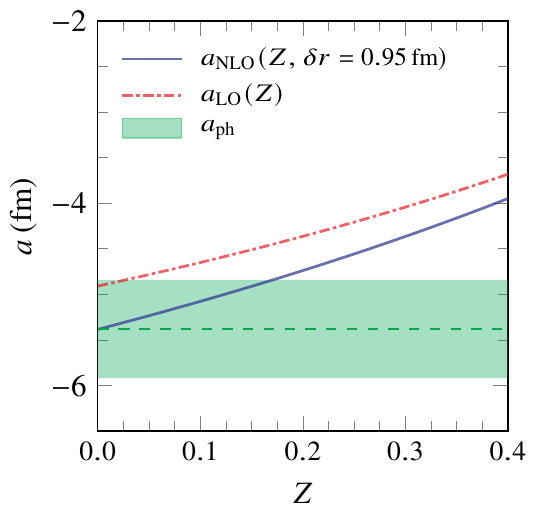}
\includegraphics[scale=0.99]{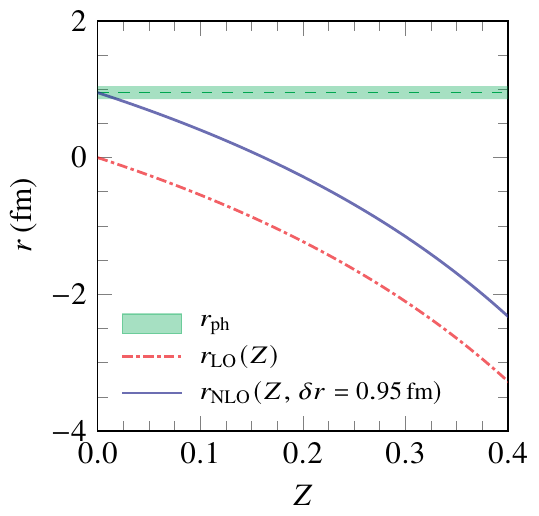}
\caption{Left: Distribution of Eq.~\eqref{eq:liHood} for different values of the probability $Z$ of the $\Tcc$ state ending in a bare elementary particle (compact tetraquark), as a function of the NLO contribution $\delta r$ [Eq.~\eqref{eq:rNLO}] to the effective range of the isoscalar $D\bar{D}^\ast$ interaction. Middle and right: $Z$ dependence of $a$ (middle) and $r$ (right) at LO (red, dashed-dotted lines) and NLO (blue, solid lines), compared with the phenomenological values (green bands) discussed in Sec.~\ref{sec:Tcc}.\label{fig:Tcc}}
\end{figure*}

\section{\boldmath An estimation of $\Tcc$ compositeness} \label{sec:Tcc}

The $\Tcc$ state has been recently discovered by the LHCb collaboration \cite{LHCb:2021auc,LHCb:2021vvq} as a prominent peak in the $D^0 D^0 \pi^+$ spectrum. It is very close to the $D^{\ast+}D^0$ threshold, since the experimental analysis throw $\Delta M_{\Tcc} = M_{\Tcc} - m_{D^{\ast+}} - m_{D^0} = -273(61)\,\text{keV}$~\cite{LHCb:2021auc} or $\Delta M_{\Tcc} = -360(40)\,\text{keV}$~\cite{LHCb:2021vvq}.\footnote{Here we do not discuss the width, the importance of which has been highlighted in the literature, \textit{e.g.} in Refs.~\cite{Yan:2021wdl,Meng:2021jnw,Fleming:2021wmk,Ling:2021bir,Lin:2022wmj}.} However, the $D^{\ast0}D^+$ threshold is only $1.4\,\text{MeV}$ above, and therefore one must consider coupled channels in order to have a fully accurate description of the state. In this situation one cannot straightforwardly apply Weinberg's compositeness criteria, which requires the coupling of the bound state to a single channel. 

The approach should be modified in the presence of close coupled channels that play an important role on the long-distance dynamics of the state, and as a consequence, they significantly modify the effective range parameter~\cite{Matuschek:2020gqe}. In Ref.~\cite{Baru:2021ldu}, it is shown for the $\chi_{c1}(3872)$ and $T_{cc}^+$ exotic states that the appearance of a large and negative effective range, which in the one-channel case would indicate the dominance of a compact component~\cite{Weinberg:1965zz}, can be naturally generated by the coupled-channel dynamics. Hence, in presence of coupled channels, and before analyzing  the estimator of the compositeness proposed in this work, it would be necessary to correct the effective range by a term~\cite{Matuschek:2020gqe,Baru:2021ldu} that stems from coupled-channel effects, and which clearly needs to be attributed to the molecular component of the state. Other efforts beyond the one just mentioned have been devoted to extend the compositeness condition to coupled channels (see references in Sec.~\ref{sec:intro}). 

However, in this work we adopt a different perspective and for illustrative purposes, we have qualitatively applied our generalization of Weinberg compositeness condition to the $\Tcc$ state within a simplified scenario. We have reduced the coupled-channel problem to a single-channel one by considering the model of Ref.~\cite{Albaladejo:2021vln} in the exact isospin limit, which we briefly outline below. The latter work performs an S-wave $D^{\ast+}D^0$, $D^{\ast0}D^+$ coupled-channel analysis in terms of two coupling constants $C_{0,1}$ and an ultraviolet cutoff $\Lambda$. If one takes common masses for the $D^{(\ast)}$ mesons, $m_{D^{(\ast)}} = (m_{D^{(\ast)+}} + m_{D^{(\ast)0}} )/2$, then the two-channel problem diagonalizes and one ends up with two independent amplitudes for each of the definite-isospin  ($I=0$ and $I=1$) sectors. Taking this limit, and using the values for the constant $C_0$ fitted in Ref.~\cite{Albaladejo:2021vln}, we  obtain $a_\text{ph} = -5.38(30)\,\text{fm}$ and $r_\text{ph}=0.95(32)\,\text{fm}$ for
the isoscalar\footnote{We assume that the $\Tcc$ state is mostly an isoscalar state, as suggested by additional experimental information besides the $D^0 D^0 \pi^+$ spectrum, see Refs.~\cite{LHCb:2021auc,LHCb:2021vvq,Albaladejo:2021vln}.} scattering length and effective range, respectively. The central values account for the averages of Eqs.~(3.9a) and~(3.9b) in Ref.~\cite{Albaladejo:2021vln}, computed for two different values of the cutoff $\Lambda=0.5\,\text{GeV}$ and $1.0\,\text{GeV}$. The quoted errors represent the addition in quadrature of the statistical error and half of the dispersion between both determinations. In this isospin limit, the $\Tcc$ binding energy (respect to the average threshold) increases, and we find $\Delta M_{\Tcc}^\text{ph} = -833(67)\,\text{keV}$ and $-856(53)\,\text{keV}$ for $\Lambda=0.5\,\text{GeV}$ and $\Lambda=1.0\,\text{GeV}$, respectively, which lead to a binding momentum $\gamma_b^\text{ph} = 40.2(1.7)\,\text{MeV}$. Taking the scale $\beta \simeq m_\pi$, one obtains $(\gamma_b^\text{ph}/\beta)^2 \simeq 8\%$, which is of the order of the uncertainty in the scattering length and smaller than that of the effective range.

For the $\Tcc$  and contrary to the case of the deuteron, we are not determining the binding momentum, the scattering length and the effective range from an experimental analysis, but instead from a phenomenological model \cite{Albaladejo:2021vln}. For this reason, we have denoted these quantities as $\gamma_b^\text{ph}, a_\text{ph}$ and $r_\text{ph}$, respectively. It is worth pointing out that two recent LQCD simulations \cite{Padmanath:2022cvl,Chen:2022vpo} have also computed the scattering length and effective range for $S$-wave $I=0$ $D\bar{D}^\ast$ scattering, and have found a value $r_0 \simeq 1\,\text{fm}$, similar to the one obtained in Ref.~\cite{Albaladejo:2021vln}, to be used in our work. We also note that, while here $r_\text{ph}$ is small and positive, the experimental value for the effective range obtained in Ref.~\cite{LHCb:2021vvq} is large and negative. However, in the latter work, the negative value is a built-in property of the model used in the analysis. More importantly, it is obtained in a coupled channel scheme, whereas $r_\text{ph}$ obtained here refers to the single channel case in the exact isospin limit. Therefore, the comparison between the two values is not  meaningful. Further discussions on this subject can be found in Refs.~\cite{Matuschek:2020gqe,Du:2021zzh,Baru:2021ldu}. 

Nevertheless, we should mention that the results of Ref.~\cite{Albaladejo:2021vln} are consistent with those found in the state of the art work of Ref.~\cite{Du:2021zzh}, where   i) the $D^{\ast+}D^0$, $D^{\ast0}D^+$ coupled-channel dynamics, ii) long-range interactions provided by the one pion exchange mechanism and iii) effects from the three-body $DD\pi$ thresholds, which lie very close to and below the two-body $D^\ast D$ ones, are accurately taken into account. Indeed, the parameters of the effective range expansion from the low-energy scattering amplitude are also reliably extracted in Ref.~\cite{Du:2021zzh}, and it turns out that both scattering length and effective range are compatible, within uncertainties, with  $a_\text{ph}$ and $r_\text{ph}$ used here, and obtained from the analysis of Ref.~\cite{Albaladejo:2021vln}. This is so once the the largest contribution to the effective range, originated from isospin breaking  related to the $D^{(*)}$-meson mass differences, that is, from the coupling of  $D^{\ast+}D^0$ to the slightly higher $D^{\ast0}D^+$ channel is discounted\footnote{Such correction is so large that it leads to a positive  residual finite range ($\approx r_\text{ph} \sim 1$ fm), which is the quantity entering in the Weinberg's relations.}  (see Table IV and Eq.~(40) of Ref.~\cite{Du:2021zzh}), as mentioned in the beginning of the section. Therefore, though the results shown below for the compositeness of the $T_{cc}^+$, obtained within the model of Ref. [76] in the isospin limit, should be considered as only qualitative ones, they might be sufficiently realistic to illustrate the performance of the estimator of the compositeness proposed in this work.

Bearing all these caveats in mind, we have computed the distribution of Eq.~\eqref{eq:liHood}, using the numerical values discussed above for $a_\text{ph}$, $r_\text{ph}$ and $\gamma_b^\text{ph}$.  The results are shown in Fig.~\ref{fig:Tcc}, similarly as done for the deuteron in Fig.~\ref{fig:deuteron}. As can be seen in the leftmost panel, $\mathcal{L}(Z,\dr)<0.5$ for $Z<0.2$ with $\gamma_b^\text{ph}/\beta\sim 0.3$, and the minimum of $\mathcal{L}(Z,\dr)$ is quite compatible with $Z=0$. Indeed in the middle panel, for $\dr=0.95\,\text{fm}(=r_\text{ph})$ such that $r_\text{NLO}(Z=0,\dr=0.95\,\text{fm})=r_\text{ph}$, it can be seen that the scattering length $a_\text{NLO}$ coincides with $a_\text{ph}$ at $Z=0$. This analysis supports a molecular picture for the $\Tcc$ state, as previous works have also concluded \cite{Feijoo:2021ppq,Du:2021zzh}, and in particular  the model of Ref.~\cite{Albaladejo:2021vln}, whose predictions for the binding momentum, scattering length and effective range  have been used here.

 \section{Summary and discussion}\label{sec:summary}
 
 We have discussed a model-independent estimator of the likelihood of the compositeness of a shallow S-wave bound or virtual state. It relies only on the proximity of the energy of the state  to the two-hadron threshold to which it significantly couples and on the experimental scattering length and effective range low energy parameters. The approach is based on NLO Weinberg's relations and it is self-consistent as long as the obtained $Z$ is small of the order of ${\cal O}(\gamma_b/\beta)$. To systematically study systems where $Z$ could take larger values, the order ${\cal O}(k^4)$ or higher terms in the ERE  would be required. This is because in those cases, it would be necessary to include additional details  to further constrain  the short-range structure of the wave-function. We have analysed the case of the deuteron (Sec.~\ref{sec:deu}), the exotic $\Dsz$ resonance (Sec.~\ref{sec:ds0}), and the $^1{\rm S}_0$ nucleon-nucleon virtual state (Sec.~\ref{sec:vir}), and found strong support to the molecular interpretation in all cases. Nevertheless, results are less conclusive for the $\Dsz$  due to the large size of the power-counting parameter $\gamma_b/\beta \sim 0.6$ and therefore the NLO approach employed here is at the limit of its applicability. 
 
 As discussed in Sec.~\ref{sec:Tcc}, the Weinberg compositeness criteria as well as the extension presented here are only valid for the case of a single channel, and the approach needs to be modified when coupled channels are present \cite{Matuschek:2020gqe,Baru:2021ldu}. To avoid this problem, in Sec.~\ref{sec:Tcc} we have discussed the case of the $\Tcc$, by reducing the coupled-channel ($D^{\ast+}D^0$-$D^{\ast0}D^+$) problem to a single-channel one in the isospin limit using the phenomenological model of Ref.~\cite{Albaladejo:2021vln}. Applying the generalization of Weinberg's compositeness condition proposed in the present work, we have shown that the parameters are compatible with a molecular interpretation of the $\Tcc$ state. These results are, however, only qualitative  and for illustrating purposes, since we have not considered the complexity of the very close thresholds $D^0D^{\ast +}$ and $D^+D^{\ast 0}$, which could make more difficult to appreciate the main ingredients of the approach (expansion) derived in this paper.

\section*{Acknowledgements}
We warmly thank E. Ruiz-Arriola for useful discussions. This research has been supported  by the Spanish Ministerio de Ciencia e Innovaci\'on (MICINN) and the European Regional Development Fund (ERDF) under contract PID2020-112777GB-I00,  the EU STRONG-2020 project under the program H2020-INFRAIA-2018-1,  grant agreement no. 824093 and by  Generalitat Valenciana under contract PROMETEO/2020/023. M.~A. is supported by Generalitat Valenciana under Grant No. CIDEGENT/2020/002. 

\bibliography{%
ModBib,
WcS}

\end{document}